\documentclass{mn2e}

\makeatletter
\let\oldcr=\@tabularcr
\makeatother

\input epsf

\usepackage{graphicx}
\usepackage{txfonts}    
\usepackage{epsfig}

\title[solarFLAG hare and hounds] {solarFLAG\thanks{solarFLAG URL:
http://bison.ph.bham.ac.uk/\,\~\,wjc/Research/FLAG.html} hare and
hounds: on the extraction of rotational p-mode splittings from
seismic, Sun-as-a-star data}

\author[Chaplin et al.]{W.~J.~Chaplin,$^{1}$\thanks{e-mail:
wjc@bison.ph.bham.ac.uk}, T.~Appourchaux$^2$, F.~Baudin$^2$,
P.~Boumier$^2$, Y.~Elsworth$^1$,\newauthor S.~T.~Fletcher$^1$,
E.~Fossat$^3$, R.~A.~Garc\'\i a$^4$, G.~R.~Isaak$^1$\thanks{George
Isaak passed away on June 5th, 2005, prior to the completion of this
work. He is greatly missed by us all.}, A.~Jim\'enez$^5$,\newauthor
S.~J.~Jim\'enez-Reyes$^5$, M.~Lazrek$^{6,3,12}$,
J.~W.~Leibacher$^{7}$, J.~Lochard$^2$, R.~New$^8$,\newauthor
P.~Pall\'e$^5$, C.~R\'egulo$^{5,9}$, D.~Salabert$^{10}$,
N.~Seghouani$^{11}$, T.~Toutain$^{1,12,13}$ \newauthor and
R.~Wachter$^{14}$\\ $^1$School of Physics and Astronomy, University of
Birmingham, Edgbaston, Birmingham, B15 2TT, U.K.\\ $^2$ Institut
d'Astrophysique Spatiale (IAS), Batiment 121, F-91405, Orsay, France;
Universit\'e Paris-Sud 11 and CNRS (UMR 8617\\ $^3$ LUAN, UMR 6525,
Universit\'e de Nice-Sophia Antipolis, 06108 Nice Cedex 2, France\\
$^4$ Service d'Astrophysique, CEA/DSM/DAPNIA, CE Saclay, 91191 Gif sur
Yvette, France \\ $^5$ Instituto de astrof\'\i sica de Canarias,
38205, La Laguna, Tenerife, Spain \\ $^6$ LPHEA, Faculte des Sciences
Semlalia, Universite Cadi Ayyad, Marrakech, Morocco\\ $^7$National
Solar Observatory, 950 North Cherry Avenue, Tucson, AZ 85719\\ $^8$
Faculty of Arts, Computing, Engineering and Sciences, Sheffield Hallam
University, Sheffield S1 1WB\\ $^9$ Dpto. de Astrofísica, Universidad
de La Laguna, La Laguna, 38206, Tenerife, Spain\\ $^{10}$ High
Altitude Observatory, National Center for Atmospheric Research,
P.O. Box 3000, Boulder, CO 80307-3000\\ $^{11}$ CRAAG, BP63,
Bouzareah, Alger, Algeria\\ $^{12}$ Laboratoire Cassiop\'ee, UMR
OCA-CNRS 6202, Observatoire de la C\^ote d'Azur, F-06304 Nice,
France\\ $^{13}$ Institute of Theoretical Astrophysics, University of
Oslo, P.O. Box 1029, N-0315 Oslo, Norway\\ $^{14}$ W. W. Hansen
Experimental Physics Laboratory, Stanford University, Stanford, CA
94305-4085\\}

\begin{document}

\maketitle

\begin{abstract}

We report on results from the first solar Fitting at Low-Angular
degree Group (solar FLAG) hare-and-hounds exercise.  The group is
concerned with the development of methods for extracting the
parameters of low-$l$ solar p mode data (`peak bagging'), collected by
Sun-as-a-star observations.  Accurate and precise estimation of the
fundamental parameters of the p modes is a vital pre-requisite of all
subsequent studies. Nine members of the FLAG (the `hounds')
fitted an artificial 3456-d dataset. The dataset was made by the
`hare' (WJC) to simulate full-disc Doppler velocity observations of
the Sun. The rotational frequency splittings of the $l=1$, 2 and 3
modes were the first parameter estimates chosen for
scrutiny. Significant differences were uncovered at $l=2$ and 3
between the fitted splittings of the hounds. Evidence is presented
that suggests this unwanted bias had its origins in several effects.
The most important came from the different way in which the hounds
modeled the visibility ratio of the different rotationally split
components. Our results suggest that accurate modelling of the ratios
is vital to avoid the introduction of significant bias in the
estimated splittings. This is of importance not only for studies of
the Sun, but also of the solar analogues that will targets for
asteroseismic campaigns.

\end{abstract}

\begin{keywords}

Sun: helioseismology -- Sun: rotation -- Sun: interior -- methods:
data analysis

\end{keywords}

\section{INTRODUCTION}
\label{sec:intro}

The solar Fitting at Low-Angular degree Group -- solar FLAG -- is an
international collaboration of helioseismologists. Its basic aims are
to actively develop new, and refine existing, techniques for analysis
of low-angular degree (low $l$) solar p mode data collected by
unresolved, Sun-as-a-star observations. Observations of this type take
a weighted average, over the visible disc, of either Doppler velocity
or intensity perturbations. This means they are very sensitive to
perturbations from the low-$l$ modes.

There are currently four active observational Sun-as-a-star programs:
the ground-based Birmingham Solar-Oscillations Network (BiSON)
[$\simeq 30$-yr database, including the Mark~I instrument on Iza\~na];
and the Global Oscillations at Low-Frequency (GOLF) and VIRGO/SPM
instruments on board the ESA/NASA SOHO spacecraft [both $\simeq 10$-yr
database]. A full 11-yr activity cycle of observations is also
available to the helioseismology community from the recent officially
completed ground-based IRIS program (although two of the sites
continue to collect data).

The unprecedented levels of S/N and resolution achieved in these long
datasets are demanding that a fresh look be taken at how the basic
parameters of the modes (e.g., frequency, frequency splitting, power,
damping etc.) are extracted. When conflicting results are given from
analyses made by different groups on different datasets, it is hard to
know from where the disagreement originates. It may have a contribution
from differences in the analysis codes, or the slightly different way
in which each instrument measures the signatures of the p modes.  A
highly coordinated approach, like FLAG, in which different codes are
tested on common datasets, offers a constructive way forward. Similar
approaches are extant elsewhere in helioseismology---for example in
the domain of local-area seismology (The Local Helioseismology
Comparison group,
LoHCo\footnote{http://gong.nso.edu/science/meetings/lohco/}).

Helioseismic parameter extraction is usually accomplished by fitting
the resonant structure of the modes, in a power spectrum of the
observations, to complicated multi-parameter models, which seek to
represent, as accurately as possible, the underlying structure
present. Analysis of this type is usually grouped under the catch-all
heading of `peak bagging'. Refinement of the Sun-as-a-star
peak-bagging is the main concern of FLAG. Those improvements to
technique and application that come from FLAG will of course also have
relevance for analysis of asteroseismic time series of other Sun-like
stars (Bedding \& Kjeldsen 2003). We envisage including as part of
FLAG's activities investigations targeted specifically to
asteroseismic problems. Put another way, the group has (implicitly)
\textit{helio}-FLAG and \textit{astero}-FLAG components to it.

Activity in FLAG has centered on a hare-and-hounds exercise.
Artificial Sun-as-a-star data, spanning 3456 simulated days -- a
length roughly commensurate to one 11-year cycle of solar activity --
were generated by WJC (the `hare'), with a S/N per unit time
characteristic of that in the GOLF and BiSON Doppler velocity data
(Chaplin et al. 2003a). These FLAG data were then fitted by nine
group members (the `hounds'). A-priori information, of the sort
available for real observations, was passed to the hounds, but nothing
more. The hounds were then asked to return to the hare, for subsequent
collation and comparison, estimates of the underlying parameters of
the low-degree modes. Estimates of the rotational frequency splittings
were the first parameters selected for scrutiny, and are the subject
of this paper. It is worth adding that a careful eye was kept on the
other parameters during analysis to check for possible complications
from cross-talk.

The layout of the rest of the paper is as follows. In
Section~\ref{sec:data} we deal briefly with how FLAG data were
constructed. In Section~\ref{sec:hounds} we summarize commonalities
in, and differences between, the strategies for parameter fitting
adopted by the hounds. Results are discussed at length in
Sections~\ref{sec:res} and~\ref{sec:diffs}. We pay particular
attention to the effects on the fitted splittings of the modeling of
multiplets containing at least three observable components; the ranges
in frequency over which the low-$l$ pairs were fitted; and a-priori
choices made regarding modeling of peak width.  Section~\ref{sec:disc}
summarizes the results, noting implications for future analyses of
helioseismic and asteroseismic solar-analogue data.

 \section{FLAG dataset}
 \label{sec:data}

FLAG datasets were constructed component by component in the time
domain. Datasets were made to contain simulated low-$l$ modes in the
ranges $0 \le l \le 5$ and $1000 \le \nu \le 5500\,\rm \mu Hz$. A
database of mode frequency, power and line width estimates, obtained
from analyses of GOLF and BiSON data, was used to guide the choice of
input values for time-series construction.

The Laplace transform solution of the equation of a forced, damped
harmonic oscillator was used to generate each component at a 40-s
cadence in the time domain, in the manner described by Chaplin et
al. (1997). Components were re-excited independently at each sample
with small `kicks' drawn from a Gaussian distribution. The model
gave rise to peaks in the frequency power spectrum whose underlying
shapes were Lorentzian. It should be pointed out that the observed
low-$l$ peaks are slightly asymmetric in shape (albeit at the level
of only a few per cent at most).

The hypothetical FLAG instrument was assumed to make its observations
from a location in, or close to, the ecliptic plane.  This is the
perspective from which BiSON (ground-based network) and GOLF (on the
SOHO spacecraft at L1) view the Sun. The rotation axis of the star is
then always nearly perpendicular to the line-of-sight direction, and
only a subset of the $2l+1$ components of the non-radial modes are
clearly visible: those having even $l+m$, where $m$ is the azimuthal
order. These components were represented explicitly in the complete
FLAG time series. The visibility for given $m$ and $l$ also depends,
though to a lesser extent, on the spatial filter of the instrument
(e.g., Christensen-Dalsgaard 1989; see also brief discussion in
Section~\ref{sec:hounds} below).  Here, we adopted BiSON-like
visibility ratios (see Chaplin et al. 2001).

The predominant contribution to the frequency splitting of the p modes
arises from the internal rotation of the Sun.  This effect can be
parameterized as an odd function of the azimuthal order, $m$, and
therefore acts to lift the frequency degeneracy in $l$ to give a
symmetric pattern in which the synodic separation between adjacent
azimuthal orders (i.e., for $|\Delta m|=1$) is $0.4\,\rm \mu Hz$.
Other agents, such as magnetic fields, that are not sensitive to the
sign of $m$ (i.e., cannot distinguish east from west) can also
contribute to the splitting (e.g., Goossens 1972), but they do so in
such a way as to introduce asymmetries in the arrangement of the
components within each multiplet (even functions of $m$)

It is at times when the level of activity on the Sun is low that the
splittings of the low-$l$ modes are very nearly symmetric. At moderate
to high levels of activity, the peaks are instead distributed unevenly
in frequency. This is because the solar-cycle changes to the
frequencies depend on both $l$ and $m$, due to the spatially
non-homogeneous, and time-dependent, nature of the driver for the
changes: the near-surface activity. For the Sun-as-a-star data,
asymmetry only becomes apparent in the frequency spacings for $l \ge
2$, when modes have more then two visible peaks. The asymmetry varies
as the solar activity cycle waxes and wanes.

In the first FLAG time series, we imposed a symmetric frequency
splitting of $0.4\,\rm \mu Hz$ on the non-radial modes (i.e., a
spacing of $0.8\,\rm \mu Hz$ between adjacent visible components in
this Sun-as-a-star dataset).  The splitting value was the same for all
modes, irrespective of overtone number and angular degree, and was
chosen to match the observed average (synodic) value extracted from
real data (e.g., see Garc\'\i a et al. 2004).

The second FLAG dataset was made to simulate the effects of the
activity cycle.  We introduced systematic changes in mode frequency
having an appropriate underlying ($l,m$) and frequency dependence. The
frequency spacings of the non-radial modes, as seen in the power
spectrum of the full time series, were therefore not
symmetric. However, the resulting asymmetry in the frequency spacings,
$\nu_{l,|m|=l} - \nu_{l,|m|=l-2}$, was smaller than the simulated
contribution of the rotation, with an average value of about $0.2\,\rm
\mu Hz$ (e.g., Chaplin et al. 2003b).

The frequency changes were introduced into the modes in the second
dataset via appropriately scaled variation of the oscillator
characteristics of each component (see, for example, Chaplin et al.
2004a). A spherical harmonic decomposition of Kitt Peak Magnetic Index
(KPMI) data, taken over the 3456-d period beginning 1992 January, was
used to calibrate the relative sizes of the changes for each
($l,m$). This was done from full-disc averages, as a function of time,
of the harmonic components of the KPMI. Solar-cycle-like changes were
also introduced in the second set in the component damping
constants. These affect not only line width but also the power in the
modes. Observational evidence for ($l,m$)-dependent changes to the
widths of the low-$l$ modes is currently marginal at best (e.g.,
Chaplin et al. 2004b). As such, activity-dependent changes in the FLAG
dataset were the same for all modes (regardless of frequency, $l$, or
$m$).

To complete both datasets, a background noise component was also added
in the time domain, the power of which increased at lower frequencies
(a second-order polynomial in $1/\nu$) in order to give
signal-to-background ratios commensurate with GOLF or BiSON-like
Doppler velocity data. Two artificial datasets were constructed by the
hare, each of length 3456\,d. In one, values of the underlying mode
parameters were held constant in time, while in the other some
solar-cycle-like effects were introduced.

The FLAG time series used for this paper are freely available at
\texttt{\small{http://bison.ph.bham.ac.uk/\~\,wjc/Research/dataFLAG.html}}.

 \section{Fitting strategies of the hounds}
 \label{sec:hounds}

The time series were made available, with 100-per-cent duty cycles,
for the nine hounds to fit (TA, FB, STF, RAG, SJJ-R, ML, DS, TT and
RW). A priori information was limited to: the cadence and length of
the datasets; and the calibration and format of the FLAG dataset. For
the purposes of this study we chose not to impose an observational
window function on the data (e.g., that from a ground-based
network). This allowed us to test parameter extraction under the more
`benign' conditions afforded by a continuous set of observations.

All the hounds fitted multi-parameter models to the resonant
structure in a power spectrum of the complete time series.
Differences in the adopted fitting strategies, and approaches to
coding the algorithms, had the potential to give variation in the
fitting parameters. Numerical inconsistencies from the second source
are subtle and hard to pin down. Those from the first are in
principle easier to identify since elements of each fitting strategy
can be listed and compared.

A dataset of Sun-as-a-star observations gives a single time series
whose frequency spectrum will contain many closely spaced resonant
peaks.  Parameter estimation must contend with the fact that the
various $m$ in the multiplets lie in very close frequency proximity to
one another. Suitable models, which seek to describe the
characteristics of the $m$ present, must therefore be fitted to the
components simultaneously. Furthermore, overlap between modes adjacent
in frequency is a cause for concern over much of the low-$l$ spectrum.
Each hound therefore took the approach most often used for
Sun-as-a-star data, and fitted modes in pairs ($l=2$ with neighbouring
$l=0$; and $l=3$ with $l=1$). All but one did so by isolating narrow
fitting `windows' centred on the target pairs.  Chosen window sizes
varied from 40 to $50\,\rm \mu Hz$ for the even-$l$ pairs, and 40 to
$60\,\rm \mu Hz$ for the odd.

The multi-parameter models to which the modes are fitted then usually
have in them contributions from the components of the target pair, and
an additional, flat offset to represent the pseudo-white, non-resonant
background (which varies only very slowly with frequency in the range
of interest). The models, however, then do not take into account that
small fraction of power in the vicinity of the target pair that comes
from nearby, weak $l=4$ and 5 peaks, and also the slowly-decaying
tails of the other even and odd-pair modes in the spectrum. These
extra contributions were allowed for by one of the hounds, who fitted
two pairs at a time.  Some of the hounds then went to a mode-by-mode
approach at very low frequencies (typically below about $1800\,\rm \mu
Hz$), where the narrower peak widths reduce substantially the overlap
from peaks of adjacent modes (an effect also called `mode blending').

While the majority varied parameters in their fitting models
simultaneously (until the fits converged), two hounds adopted a
multi-step iterative procedure (e.g., see Gelly et al. 2002). This
involved holding a few parameters constant during some iterations,
the intention being to reduce fitting cross-talk between strongly
correlated parameters.

Further differences were present in the detail of the numbers, and
types, of parameters the hounds sought to estimate by fitting. For
example, some attempted to fit for peak asymmetry, where individual
components were in all cases represented by the asymmetric formula of
Nigam \& Kosovichev (1998); while others, after preliminary analysis
had indicated none was present, fixed the asymmetry at zero and
submitted revised estimates from a (simpler) Lorentzian-based fitting
model. There was also a division between those hounds who constrained
the widths of all components of the $l=2$/0 or 3/1 pairs to be the
same (again, common practice at low $l$); and those who instead fitted
two widths, one for each ensemble of components in a mode. However, of
all the potential sources of discrepancy, the modeling of peak
visibility in multiplets was found to be potentially the most
significant.

When $l > 1$ a mixture of $|m|$ is observed. Since the visibility
filter of the observations is strongly $|m|$ dependent, a suitable
strategy must be adopted to allow for this. If $H(l,m,n)$ is the
height of any given component as it appears in the power spectrum
(i.e., the maximum power spectral density), the height ratio
$\epsilon(l,n)$ for $l=2$ and 3 multiplets is defined to be:
 \begin{equation}
 \epsilon(2,n)= H(2,0,n) / H(2,|2|,n),
 \end{equation}
 \begin{equation}
 \epsilon(3,n)= H(3,|1|,n) / H(3,|3|,n)
 \end{equation}
respectively. The underlying ratios used to construct the FLAG data
were $\epsilon(2,n) = 0.54$ for all $l=2$ modes, and $\epsilon(3,n) =
0.38$ for all $l=3$ modes. The values were chosen by WJC to match
those found in the BiSON data.

It is common practice to include the height ratio as some fixed
parameter in the fitting. However, if the assumed value does not match
that of the actual data, the fitted frequency splitting will be
affected. That this is a potentially serious problem for Sun-as-a-star
analysis was first discussed at length by Chaplin et al. (2001). They
showed that when the adopted value of $\epsilon(l,n)$ is larger than
the true, underlying value the splittings will tend to be
overestimated; while the splittings will tend to be underestimated
when $\epsilon(l,n)$ is too small. The bias that is introduced is more
severe at high frequency.

The hounds were given no a-priori information regarding the underlying
height ratios in the FLAG data. As noted in Section~\ref{sec:data}, in
real observations the exact ratios are determined by the spatial filter
response of the instrument. The filter will be different for intensity
and Doppler velocity observations; what is more, it will also differ
from Doppler instrument to Doppler instrument, due to, for example,
differences in the way the target Fraunhofer line is sampled to
determine the Doppler signal, and the width of Fraunhofer line itself
(Doppler imaging). With information on the nature of the observation,
it is then possible to calculate a theoretical set of height ratios
(e.g., see examples in Christensen-Dalsgaard 1989; Appourchaux et
al. 2000a). It is worth adding that making the determination as
realistic as possible is non-trivial, and as such is not without its
problems.

Although the hounds could have been given filter response information
for the hypothetical FLAG instrument, we chose instead to explore the
importance for the final results of letting the hounds make their own
decisions. All adopted the strategy of fitting with fixed height
ratios of their own choosing. These varied from $\epsilon(2,n) = 0.50$
to 0.60 at $l=2$, and from $\epsilon(3,n) = 0.19$ to 0.60 at $l=3$.

Finally, all hounds restricted fits for the splitting to frequencies
below about $4000\, \rm \mu Hz$. This was because of the known
problems of estimating the parameter at high frequencies, where there
is substantial overlap between adjacent peaks (e.g., Chaplin et
al. 1998, 2001; Appourchaux et al. 1998; 2000b). At $4000\, \rm \mu Hz$
the ratio of the separation in frequency of adjacent $m$ to the peak
width is $\simeq 0.08$. The splittings are as a result both strongly
overestimated, and poorly constrained. (See Garc\'\i a et al. 2004,
and Chaplin et al. 2004c, for discussion of the contribution of
high-frequency splittings to rotation inversions.)

\section{RESULTS}
\label{sec:res}

\begin{figure*}
 \centerline{ \epsfbox{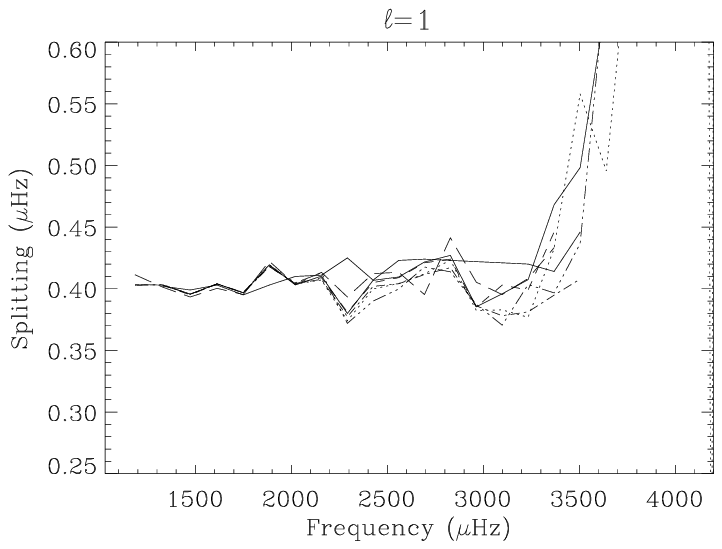} \quad \epsfbox{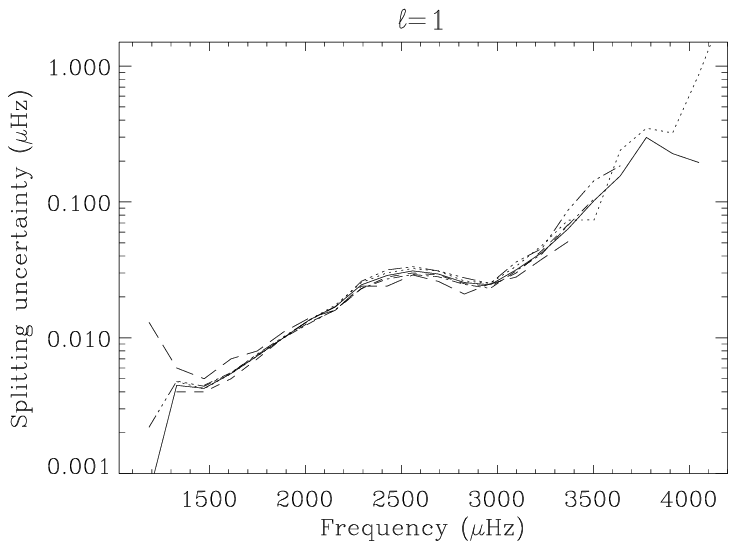}}
 \centerline{ \epsfbox{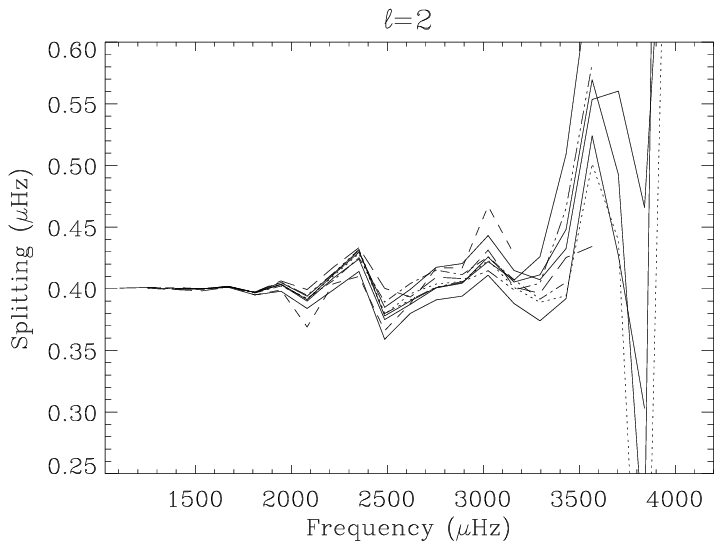} \quad \epsfbox{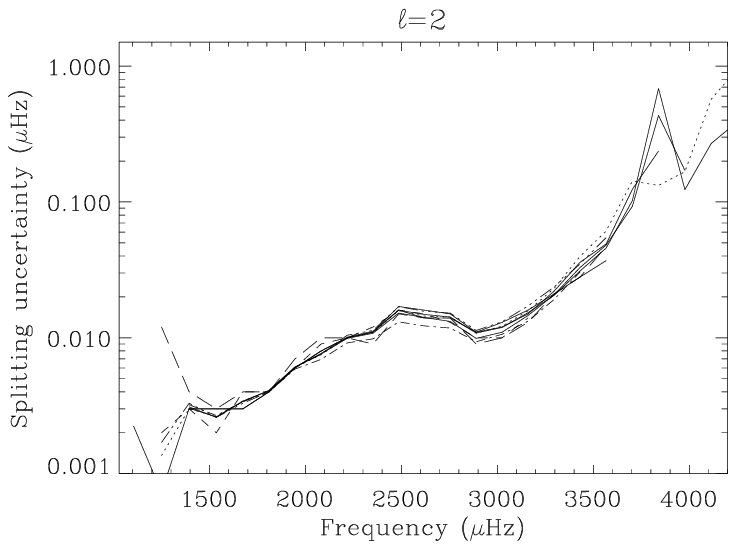}}
 \centerline{ \epsfbox{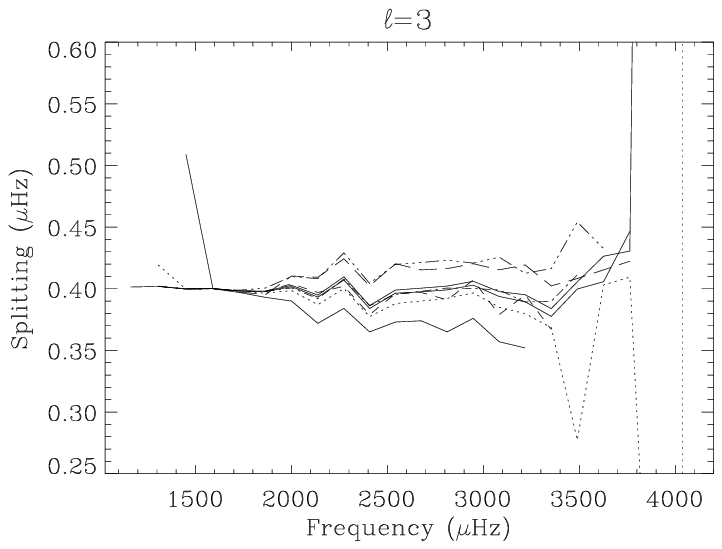} \quad \epsfbox{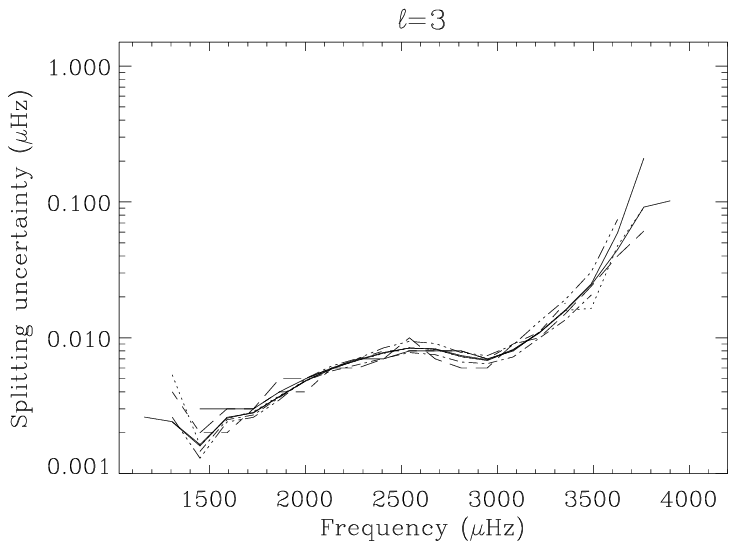}}

 \caption{Fitting results of the hounds (various curves). In the
 left-hand panels: extracted splitting estimates; and in the
 right-hand panels: the associated, formal uncertainties.}

 \label{fig:smain}
 \end{figure*}

The curves in the left-hand panels of Fig.~\ref{fig:smain} show the
rotational splitting estimates of each of the hounds, extracted from
the FLAG dataset with no temporal parameter variations. Since the
prominent outer components, with $l=|m|$, dominate the fitting (see,
for example, discussion in Chaplin et al. 2004a), the fitted
splittings are close to the sectoral mode splittings:
 \begin{equation}
 \delta\nu_s(l,n) = \frac{[\nu_{l,n,m=l} - \nu_{l,n,m=-l}]}{2l},
 \label{eq:sin}
 \end{equation}
The fitted splittings are to be compared with the input value
$\delta\nu_s(l,n)=0.4\,\rm \mu Hz$ present in all modes with $l \ge
1$. The right-hand panels show the formal uncertainties associated
with the splittings. All hounds extracted these in the same manner,
taking, for each fit, the square root of the corresponding diagonal
element of the inverted Hessian fitting matrix.

A very similar pattern of splitting results was obtained from fits to
the second, solar-cycle-modulated dataset. Here, the fitting codes had
to cope with an underlying arrangement of $m$ peaks at $l \ge 2$ that
was slightly asymmetric in frequency.  It is possible to extract the
multiplet asymmetry from long Sun-as-a-star datasets (e.g., Thiery et
al. 2001; Chaplin et al. 2003b, 2004a; Garc\'ia et al. 2004); and, by
implication, the artificial FLAG datasets here.  However, the
estimates are poorly constrained because the asymmetry is typically
smaller in size than $\delta\nu_s(l,n)$; and the inner components
[having ($l,|m|$) of ($2,0$) and ($3,1$)] are less prominent, and
harder to distinguish, than the outer sectoral peaks.  It should come
as no surprise that since the outer $m$ dominate the fitting the
returned splittings are hardly affected by choosing to fit to a model
that takes no account of the multiplet asymmetry (the type of model
fitted by the hounds). An added benefit is that, with fewer parameters
to fit, the extracted splittings are also better constrained. Because
of the similarity of the results from the two datasets, we show in
the remainder of the paper results obtained from the first,
`stationary' FLAG dataset only.



 \begin{figure*}
 \centerline{ \epsfbox{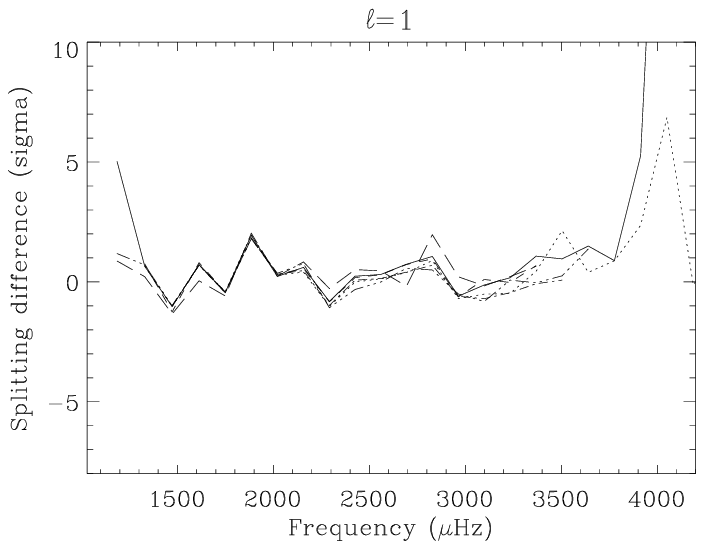} \quad \epsfbox{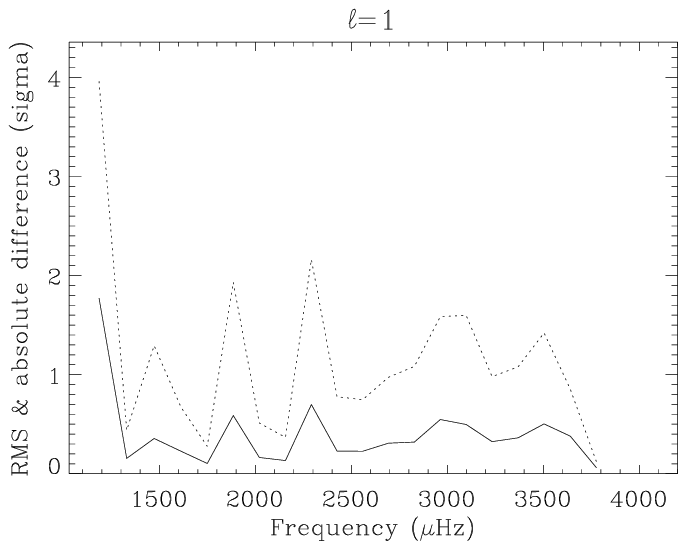}}
 \centerline{ \epsfbox{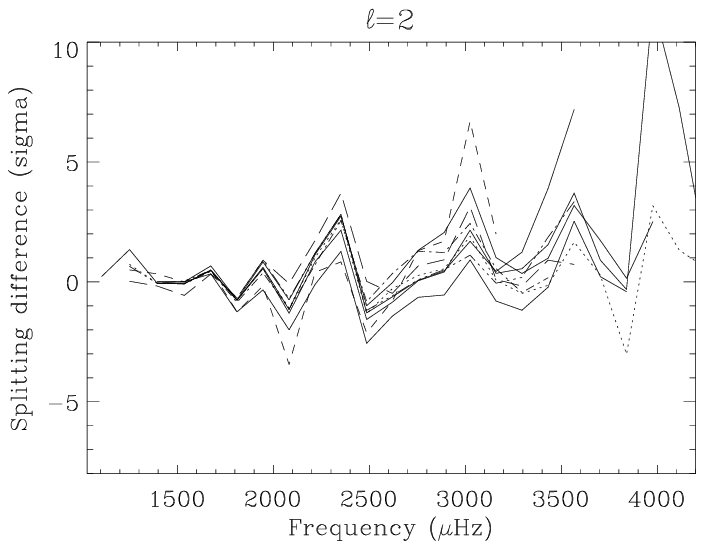} \quad \epsfbox{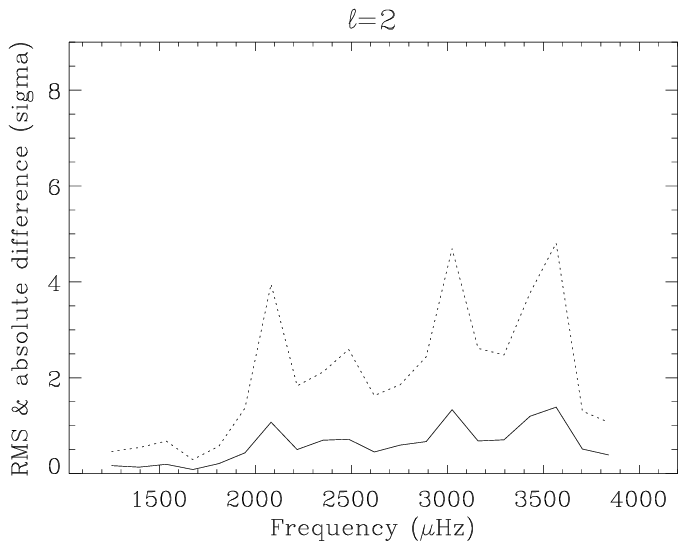}}
 \centerline{ \epsfbox{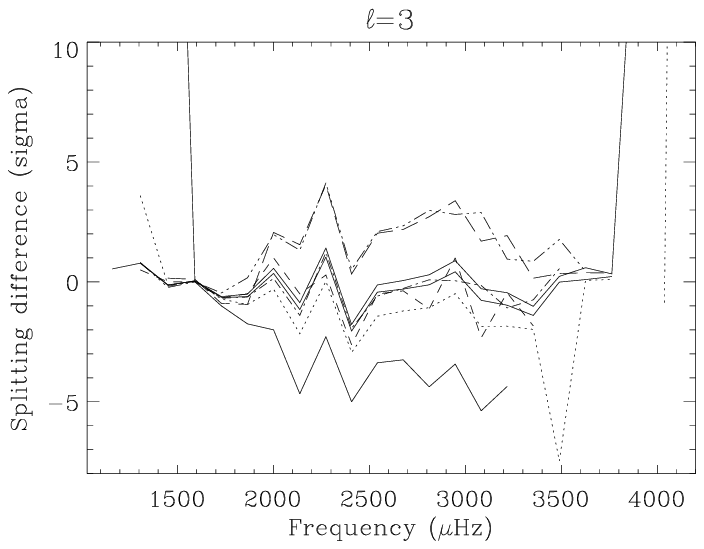} \quad \epsfbox{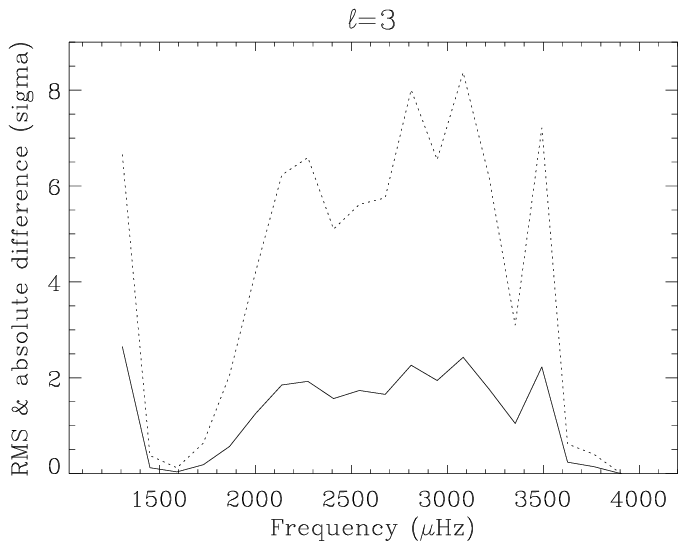}}

 \caption{Left-hand panels: Fitted minus input splittings for each of
 the hounds, divided by mean fitted uncertainty. Right-hand panels:
 \textsc{rms} (solid curve) and extreme, absolute deviations (dotted
 curve) of the residual difference curves in the left-hand panels.}

 \label{fig:srms}
 \end{figure*}


As seen in Fig.~\ref{fig:smain}, the hounds' results for $l=1$ were
quite accurate over most of the fitting range. Moreover, the different
curves shared many features. This revealed the extent to which fits
were influenced strongly by the common realization noise, in spite of
differences in fitting strategies and execution. WJC made fits to
further independent realizations of the stationary FLAG dataset, which
confirmed the features did not have their origin in another effect. It
is worth emphasizing that while `features' from the realization noise
are common between hounds, they are nevertheless unpredictable from
dataset to dataset. The $l=2$ and 3 curves also had common features
resulting from the noise.

The now familiar overestimation of the Sun-as-a-star splittings was
present at higher frequencies in the hounds' data (e.g., Chaplin et
al. 2001). This resulted in mode blending, and was most pronounced at
$l=1$, since at this degree the sectoral components lie in closer
proximity than at $l=2$ and 3.

Let us now turn our attention to the differences between the
hounds. To determine the significance of these differences we
subtracted the input splitting of $0.4\,\rm \mu Hz$ from all fitted
estimates.  Each splitting residual was then divided by the mean
uncertainty for that mode, the uncertainty being an average across the
hounds. The calculated difference residuals -- in units of the
typical, fitted uncertainty -- are plotted in the left-hand panels of
Fig.~\ref{fig:srms} (same line styles as previous). The various curves
therefore show how the results of the hounds differed from one
another. The \textsc{rms} and absolute, extreme deviations of the
difference curves are plotted in the right-hand panels.

Each splitting has an uncertainty arising from the intrinsic
(stochastic) noise on the data, and it is the formal uncertainties
from the fitting that seek to provide an estimate of this internal
scatter. The formal uncertainties might not, however, provide an
accurate measure of the full, external precision in each
parameter. Indeed, differences in the results of the hounds represent
an additional source of uncertainty for the splittings, over and above
that arising from the intrinsic noise. This extra source of error can
be estimated from the \textsc{rms} curves in Fig.~\ref{fig:srms}.

The \textsc{rms} values at $l=1$ were typically but a small fraction
of the fitting uncertainties, indicating the fitting uncertainties
were a reasonably good indicator of the actual precision of the
splitting estimates. Analysis of fits to additional realizations of
the FLAG dataset (made by WJC) bore out the conclusion. This was
manifestly not the case at $l=2$ and 3. Here, the splitting curves of
the hounds departed noticeably from one another, although the `shapes'
did remain similar (from the common, stochastic noise).  The
corresponding \textsc{rms} values were comparable to, or larger in
size than, the internal fitting uncertainties over much of the fitting
range. At $l=3$, extreme differences for some modes exceeded the
$6\sigma$ level.

The sharp, low-frequency increase in the $l=1$ and 3 \textsc{rms}
curves merits a brief comment. The modes in the simulated datasets
were extremely long lived at these frequencies and the S/N ratios were
very low (to match the real observations).

Some poor-quality fits made in this regime, which fail to give
accurate estimates of the mode parameters, will return greatly
overestimated linewidths and large splitting uncertainties.  That
said, any one of these fits will tend not to give an out-of-line datum
on the normalised difference plot, owing to the large fitting
error. Those fits that `latch' on instead to prominent spikes from the
broadband background will, since the associated uncertainties will be
very small. The low-frequency parts of the $l=1$ and 3 curves in
Fig.~\ref{fig:srms} show the presence of occasional poor fits that are
of this second type, having the expected small splitting uncertainties
(because the fitted noise spikes are sharp). This suggests that
estimates of the frequencies and splittings of very low-frequency
modes should therefore not always be regarded as being accurate (which
is usually, implicitly, taken to be the case).

 \section{On the differences between hounds}
 \label{sec:diffs}

To understand the differences uncovered in Figs.~\ref{fig:smain}
and~\ref{fig:srms} we tested, in turn, the impact of each of the
main strategy choices adopted by the hounds.

The first issue we considered involved a simple a-priori choice:
should one constrain all components in a mode pair to have the same
fitted width? The second and third strategy options involved instead a
free choice of real-valued constraints in the fitting models. These
were: first, to fix the $m$-component height ratios; and second, to
fix the size of the fitting window.  The potential for a continuum of
unwanted bias was therefore present, with the size of bias dependent
on the values of the imposed model parameters.

One issue we did not address in detail was the impact of peak
asymmetry, and attempts to fit for that asymmetry. The mechanism for
generation (in the time domain) of the FLAG data gave peaks in the
frequency domain that were Lorentzian. However, observations of the
low-$l$ modes show the presence of small amounts of asymmetry. Our
work, in particular fitting the FLAG data to models with and without
asymmetry, raised some interesting questions regarding the impact of
the asymmetry parameter. Artificial time series data, which include
resonant asymmetry, are needed to properly address this issue, i.e.,
data for which the asymmetry characteristics may be modified on a
datum-by-datum basis. This is the subject of ongoing work, and will
be left to a future paper.

It is possible to recover fairly accurate estimates of the input
splittings from serendipitous combinations of the fitting constraints,
i.e., the effects of bias from the three targeted strategy choices may
just happen to cancel fortuitously. The optimal, and desirable,
combination is clearly one in which bias from all choices is
minimized. This cross-talk should be borne in mind in what follows.

\begin{figure*}
 \centerline{ \epsfbox{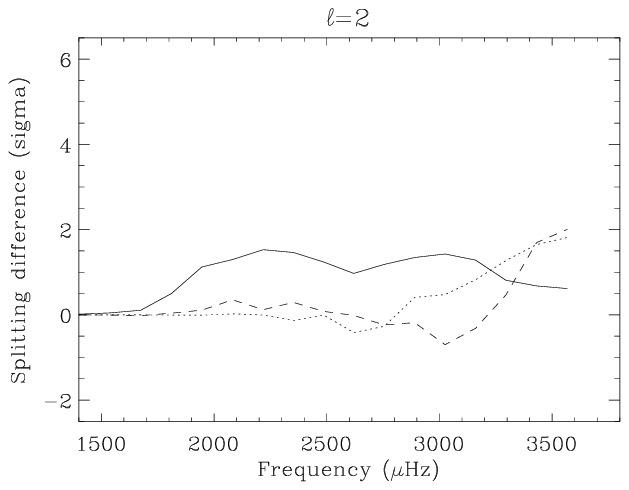} \quad \epsfbox{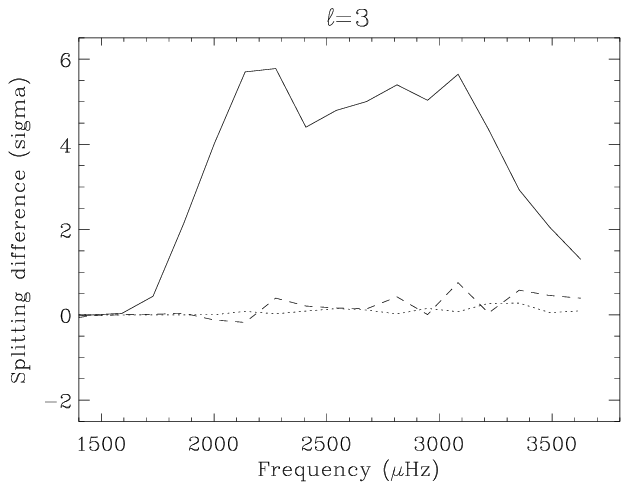}}

\caption{Impact of several effects on the splittings extracted at
$l=2$ and 3. Each line shows splitting differences, in units of the
typical uncertainty, between two strategies. Differences are between
fitting with: high and low values of $\epsilon(l,n)$, with values of
0.6 and 0.5 at $l=2$, and 0.6 and 0.2 at $l=3$ [solid line]; 50 and
40-$\rm \mu Hz$-wide windows [dotted]; and two and one width
parameters for a pair [dashed]. (See text for more details.)}

 \label{fig:probs}
 \end{figure*}

Fig.~\ref{fig:probs} gives a concise snapshot of effects
contributing to the disagreement at $l=2$ and 3, where the
significance of the differences is a potential cause for concern. To
combine these on a single plot, we show splitting
\textit{differences}, in units of the typical fitting uncertainty,
given by several \textit{changes} of fitting strategy. Rather than
use data from all hounds, we chose to use results from a single
peak-bagging code (the pair-by-pair code of WJC), with the various
fitting options activated. It is worth adding that almost identical
results were given by the pair-by-pair codes of SJJ-R and TT.



 \subsection{Modeling peak width}
 \label{sec:shape}

The dashed lines in each panel of Fig.~\ref{fig:probs} show the
(normalized) shift in splitting given by changing from fitting a
single width for all components in a pair, to two widths, one for each
mode (in the sense two-parameter minus single-width-parameter
results). The sets of parameters needed to make these curves were
returned from fits where the correct height ratios, $\epsilon(l,n)$,
were adopted, and 50-$\rm \mu Hz$-wide fitting windows were applied.
Evidently, different combinations of imposed constraints, and output
results, were possible. However, the plotted data give a reasonable
guide to the typical \textit{changes} given by changes to the width
strategy

The width strategy choice is a familiar one for those fitting
Sun-as-a-star data. Fig.~\ref{fig:probs} suggests the choice has an
insignificant impact on the $l=3$ splittings. At $l=2$, the effect is
quite modest over most of the frequency range, and only exceeds
$1\sigma$ at higher frequencies, where freeing the widths of each pair
makes the fitted splittings rise. Here, the shift is not due to
inaccurate modeling of the underlying widths: damping coefficients in
the stationary FLAG dataset were only frequency, and not ($l,m$),
dependent. Changes in width over narrow ranges in frequency were very
small; a single width was therefore the better fitting option for each
FLAG pair. Indeed, the width-strategy results suggest fitting with
constrained widths may be a good choice for real data.  However, our
tests took no account of the possible impact of dependence of the
widths on $l$ and $m$, or of a differing solar-cycle
signature. [Variation of the shifts, with $l$ and $m$, is observed in
data of medium $l$ (Komm, Howe \& Hill 2002), for which it is possible
to achieve much higher precision than at low $l$, where the evidence
for degree dependence is only tentative (Chaplin et al. 2003c).]
These are obvious avenues for future FLAG work.

 \subsection{Modeling of $m$-component height ratios}
 \label{sec:mrat}

\begin{figure*}
 \centerline{\epsfbox{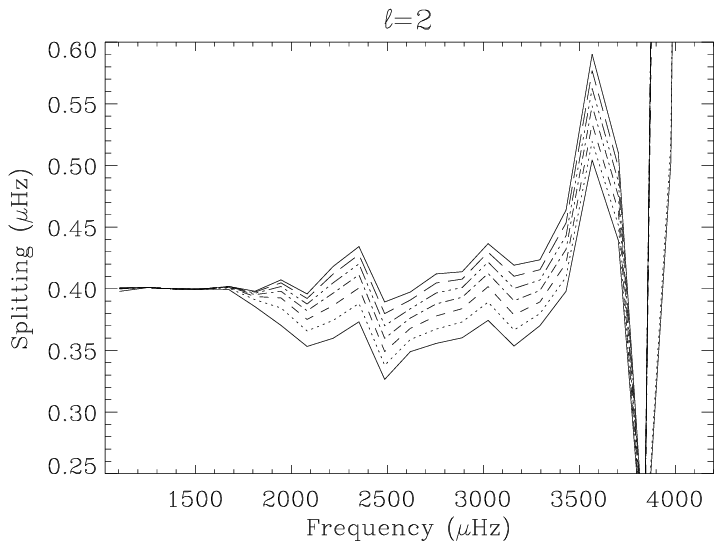}
 \quad \epsfbox{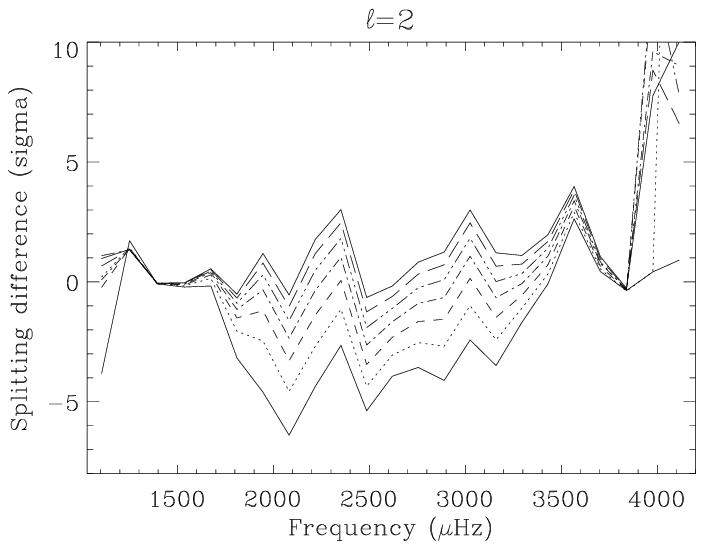}}
 \centerline{\epsfbox{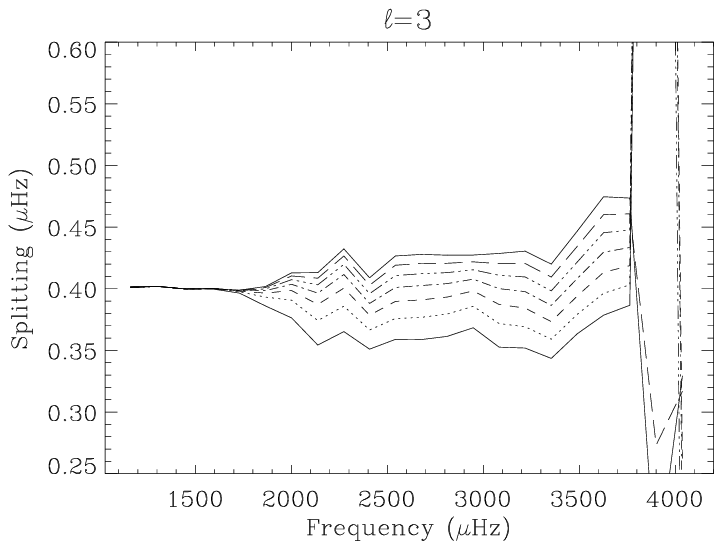}
 \quad \epsfbox{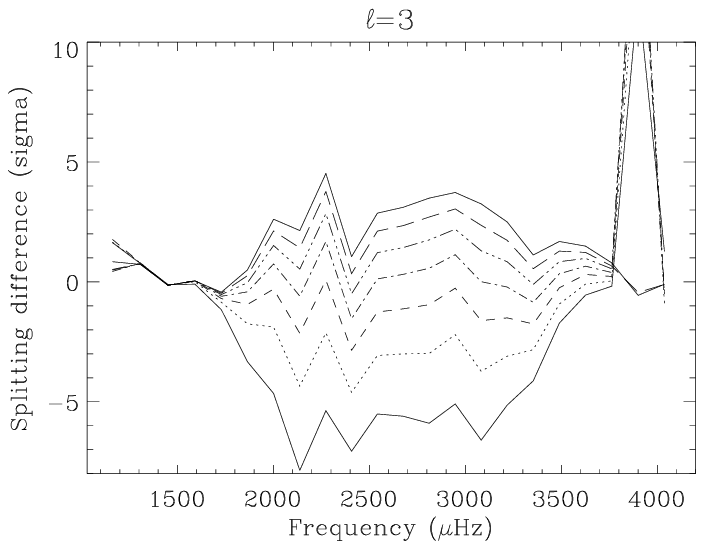}}

 \caption{Impact of fitting with different fixed $m$-component height
 ratios. The code of WJC was used to generate the results. Values
 adopted for $\epsilon(l)$ ranged from from 0.1 (giving data described
 by lower solid curve in each panel) up to 0.7 (uppermost, solid curve
 in each panel) in uniform steps of 0.1 (the intermediate curves).
 The input values were 0.55 at $l=2$ and 0.38 at $l=3$. Shown in
 the left-hand panels: the extracted splittings. Shown in the
 right-hand panels: fitted minus input splittings, divided by
 corresponding mean fitted uncertainties.}

 \label{fig:mdep}
 \end{figure*}

Each of the hounds set the height ratios, $\epsilon(l,n)$, at fixed
values prior to fitting. The solid curves in Fig.~\ref{fig:probs}
show the changes in splitting given by going from a low to high
fixed ratio (in the sense high minus low data). To make the curves,
we took $\epsilon(l,n)$ at the extreme ends of the ranges used by
the hounds. This meant values of 0.5 and 0.6 at $l=2$, and 0.2 and
0.6 at $l=3$. Clearly, a higher value of the ratio lead to a larger
fitted splitting. The point is made more fully, and the ranges of
potential bias are revealed, in Fig.~\ref{fig:mdep}. It shows
results of refitting the FLAG spectrum (again using the code of WJC)
with the height ratios at values ranging from 0.1 (splitting
estimates indicated by lower, solid line in each panel) to 0.7
(uppermost, broken line in each panel), in uniform steps of 0.1.
Similar results were obtained from an independent analysis by RAG.



The range of selected ratios, and the resulting potential for bias
in the fitted splittings, was evidently larger at $l=3$ than at
$l=2$ (as evidenced by the solid curves in Fig.~\ref{fig:probs}).
Indeed, simple inspection of the hounds' data in
Fig.~\ref{fig:smain} shows that some of the $l=3$ fits had been
strongly affected by the chosen $\epsilon(3,n)$ (for example the
uppermost two curves and the bottom curve in the lower left-hand
panel of Fig.~\ref{fig:smain}).

\begin{figure*}
 \centerline{ \epsfbox{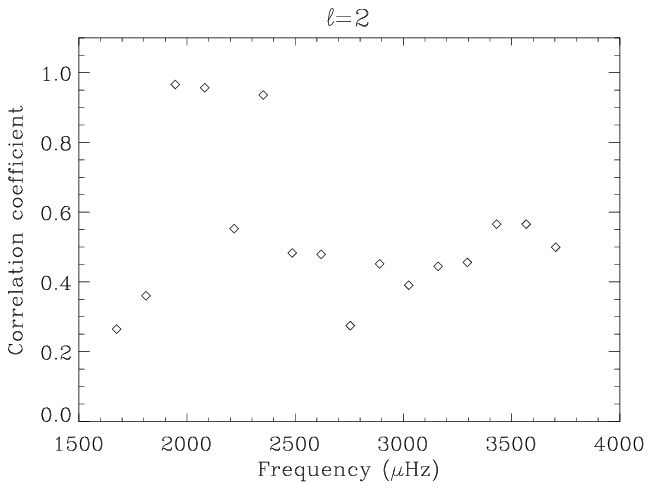} \quad \epsfbox{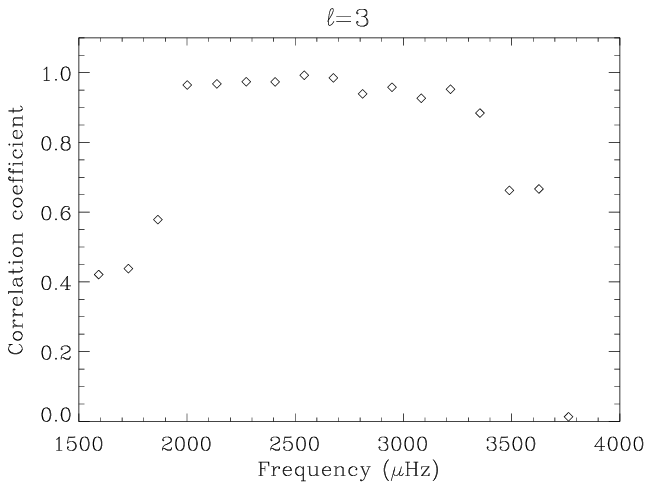}}

 \caption{Pearson correlation coefficient between the fitted splittings
 and the adopted $\epsilon(l,n)$ of the hounds.}

 \label{fig:sdm}
 \end{figure*}

The correlation between the fitted splittings of the nine hounds and
the fitting height ratios they chose is shown in Fig.~\ref{fig:sdm}.
At $l=2$ the correlation is seen to be rather marginal. In the light
of the results in Fig.~\ref{fig:probs} this was only to be expected:
changes to the splitting from the choice of $\epsilon(2,n)$, while
probably the largest source of bias in play, are not expected to be
significantly larger than the shifts arising from the other changes
of strategy. However, the expectation at $l=3$ is that height-ratio
selection will dominate over the other choices. The higher values of
correlation bear this out.


\begin{figure*}
 \centerline{ \epsfbox{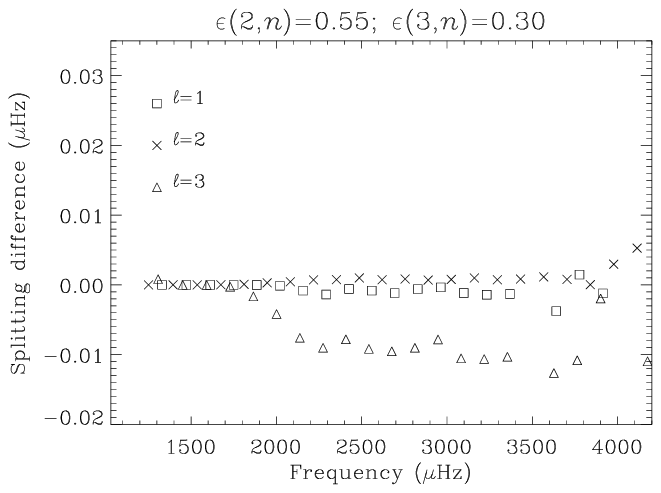} \quad \epsfbox{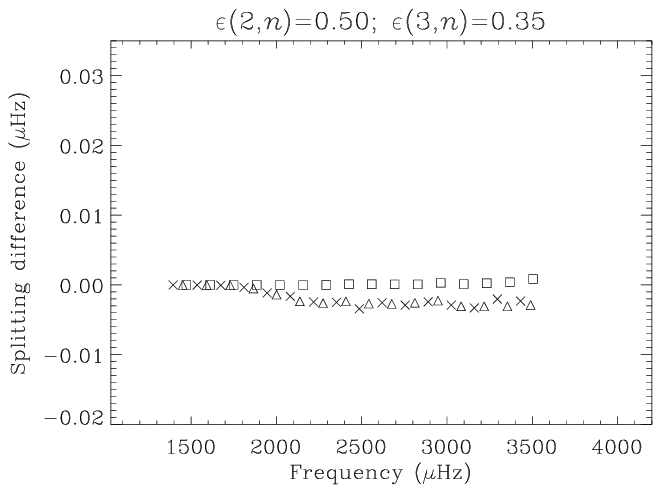}}
 \centerline{ \epsfbox{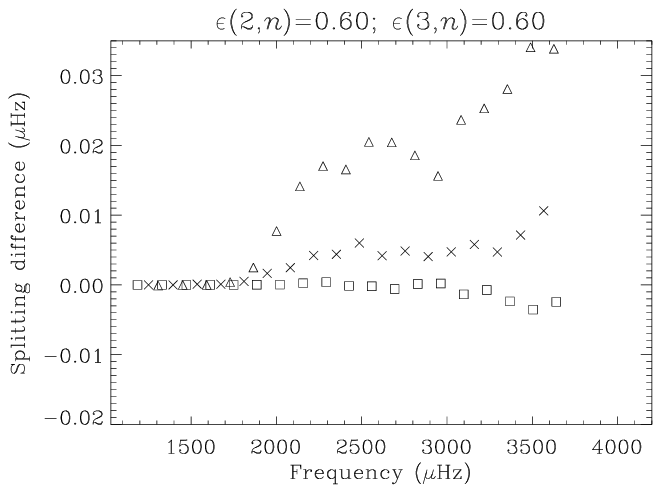}}

 \caption{Impact on extracted splittings of three of the hounds of
  re-fitting with the correct height ratios, $\epsilon(l,n)$ (0.55 at
  $l=2$ and 0.38 at $l=3$). Plotted are differences between: fitted
  splittings given with originally selected height ratios (in plot
  titles) and splittings given with the correct ratios. The $l=1$
  splittings are affected slightly by cross-talk in the fits.}

 \label{fig:meff}
 \end{figure*}

\begin{figure*}
 \centerline{ \epsfbox{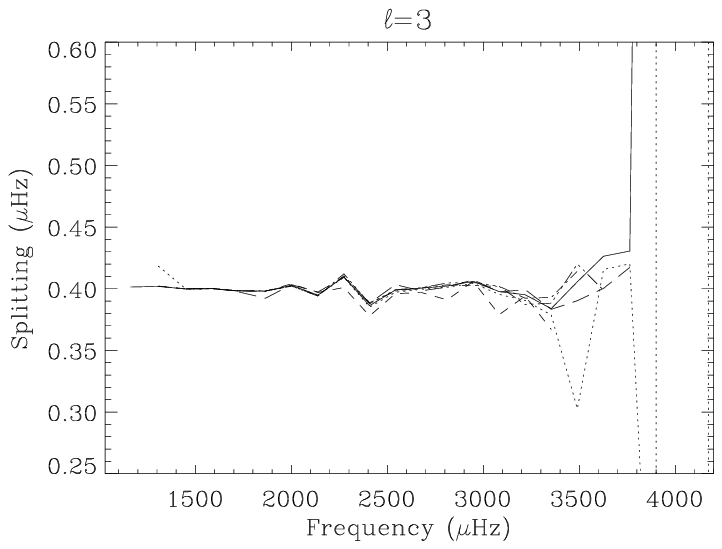} \quad \epsfbox{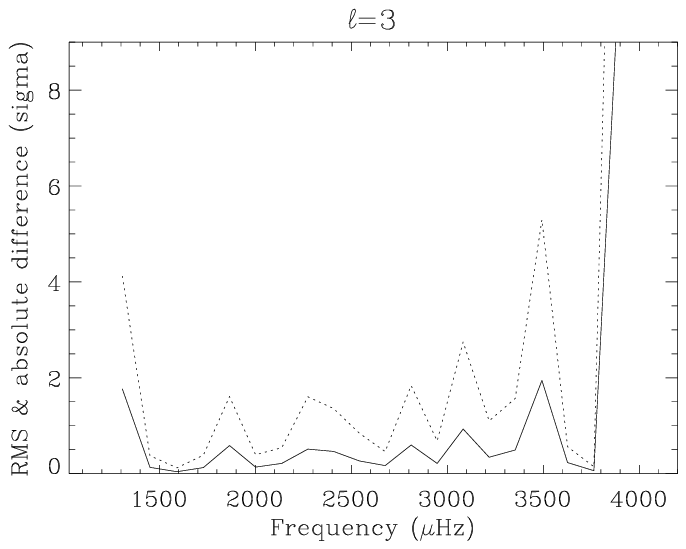}}

 \caption{Overall impact at $l=3$ of re-fitting with the correct
 height ratios. Shown in the left-hand panel: the re-fitted splittings
 of the nine hounds; and in the right-hand panel: the \textsc{rms} scatter,
 in units of the typical formal uncertainty, between hounds. These data are to be
 compared with those of the original fits in Fig.~\ref{fig:smain}
 (bottom left-hand panel) and Fig.~\ref{fig:srms} (bottom right-hand
 panel).}

 \label{fig:l3rat}
 \end{figure*}

To formally verify the impact on the fitting, the hounds were asked to
refit the FLAG data with the input height ratios in their
models. Fig.~\ref{fig:meff} shows the differences in splitting that
resulted for a sample of three of the hounds, in the sense: results
given with assumed ratios minus those given with the actual input
ratios, with the originally chosen ratios indicated in the plot
titles. While the $l=2$ splittings of all hounds were affected by the
re-analysis the changes were, as expected, not noticeably
significant. The lower panel of Fig.~\ref{fig:meff} was the most
extreme case. This resulted in there being only a modest impact on the
$l=2$ \textsc{rms} residual difference curve in Fig.~\ref{fig:srms}
(middle right-hand panel). The original choices for $\epsilon(3,n)$
had been rather more varied, and re-fitting brought significant
changes. Indeed, the scatter between hounds was all but removed at
$l=3$ (as shown in Fig.~\ref{fig:l3rat}) confirming this effect as the
dominant source of disagreement.



It is possible to fit for the height ratio. However, high intrinsic
resolution is needed to extract usable estimates of the parameter.
In the solar case, this demands time series of length approximately
one year or more. Even then, robust estimates can only be extracted
from a certain range in frequency. Furthermore, other parameters may
then also be less well constrained.

The lower-frequency modes are in principle best suited to the task,
since their narrow peaks allow for a clear separation of the
constituent $m$. However, S/N in the power spectral density of the
modes falls off rapidly at low frequency. The less prominent, inner
components may as a consequence be too weak to extract an estimate
of the ratio if modes at too low a frequency are selected for
scrutiny. At frequencies above about $\simeq 3100\, \rm \mu Hz$ the
increased severity of mode damping compromises the reliable
estimation of the underlying $\epsilon(l,n)$, since the wide peaks
overlap.

Chaplin et al. (2001) found that in the solar case a suitable range to
attempt the extraction was $1800 \la \nu \la 3000\, \rm \mu Hz$ over
which one of the pertinent parameters here -- the ratio of the
separation in frequency of adjacent $m$ to the peak width -- varies
from $\simeq 3.5$ to $\simeq 0.9$. An average of the fitted values,
over the several orders in $n$ covered by the selected frequency
range, will then give reasonably well constrained estimates of the
underlying ratios. For example, WJC obtained average estimates,
$\left< \epsilon(l) \right>$, of $\left< \epsilon(2) \right> = 0.57
\pm 0.04$ and $\left< \epsilon(3) \right> = 0.41 \pm 0.02$ from fits
to the FLAG data. Both values lie within one sigma of the input
values.

 \subsection{Choice of fitting window}
 \label{sec:win}

The dotted lines in each panel of Fig.~\ref{fig:probs} show
representative examples of the impact on the splittings of changes
to the fitting window size. Here, the plotted shifts are between
fits with 50 and 40-$\rm \mu Hz$-wide windows (in the sense
wide-window minus narrow-window data). The effect of the window size
on the pair-by-pair approach turns out to be most severe if, as is
usually the case, leakage from neighbouring $l=4$ and 5 modes is not
allowed for explicitly in the fitting model.

Analyses of Sun-as-a-star data taken in Doppler velocity concentrate,
quite rightly, on the most prominent modes present in such
data---those with $l \le 3$. However, with the advent of longer,
higher-quality Doppler velocity datasets it is now possible to extract
reasonably well-constrained estimates of some parameters of $l=4$
modes (e.g., Chaplin et al. 1996; Lazrek et al. 1997), and even some
of $l=5$ (Lazrek et al. 1997) [the latter being most conspicuous in
the near-continuous GOLF dataset]. These modes are nevertheless very
weak in comparison to their lower-$l$ counterparts. In BiSON data, the
maximum power spectral density of the most prominent $l=4$ and 5 modes
is typically $\simeq 40$ and $\simeq 300$-times lower, respectively,
than in nearby $l=0$ modes.  To ensure robustness in fits attempted on
the sextupole and decupole modes, these higher $l$ are usually given a
simplified representation in the fitting models, for example by
including only the more prominent $|m|=4$ and 5 components. The point
here is that, although these modes are weak, their presence may bias
the results of any fits that fail to account for the contribution they
make to the power in the fitting window.



 \begin{figure*}
 \centerline{ \epsfbox{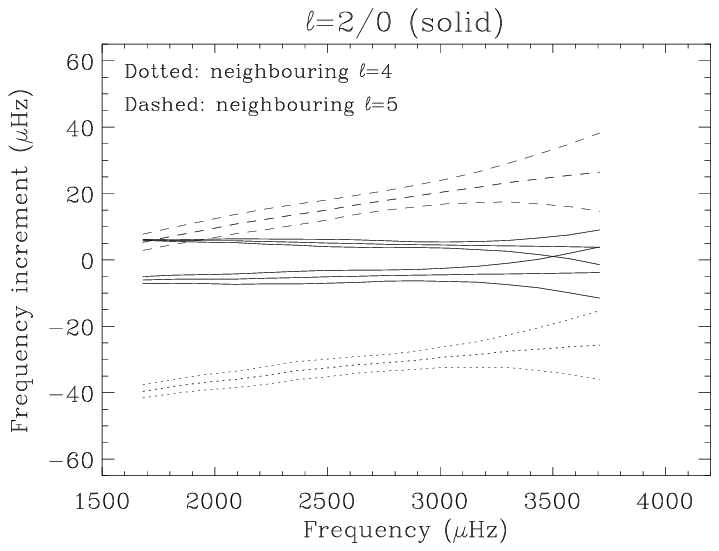} \quad \epsfbox{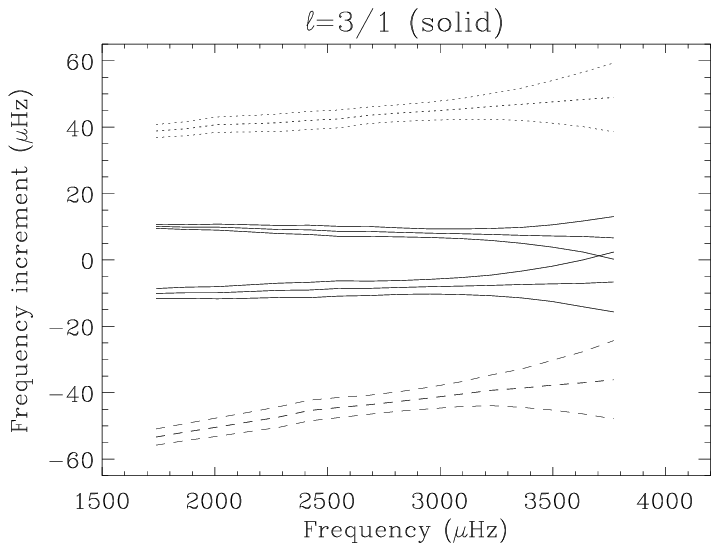}}

\caption{Location, in frequency, of $l=4$ and 5 modes with respect
to their more prominent $l=2/0$ and 3/1 counterparts. The solid
lines bound regions in frequency where the $l=0$ and 2 modes
(left-hand panel) and $l=1$ and 3 modes (right-hand panel) are most
prominent. Boundaries mark locations one linewidth of a mode (at
that frequency) beyond the location of the outer $m = \pm l$ peaks.
The dotted and dashed lines do the same, respectively, for each of
$l=4$ and 5.}

 \label{fig:where}
 \end{figure*}


The FLAG dataset includes $l=4$ and 5 modes, and, indeed, one of the
hounds allowed for their presence during fitting. The other hounds
adopted the more usual approach, and did not. The potential
influence of this on the fitting can be understood from the
placement in frequency of the $l=4$ and 5 modes with respect to the
even ($l=2/0$) and odd ($l=3/1$) pairs. This positioning is shown in
the echelle-like plots in Fig.~\ref{fig:where}. The solid lines in
the left-hand panel bound regions in frequency where the $l=0$
(upper triplet of lines) and $l=2$ modes (lower triplet) are most
prominent. The boundaries actually mark locations that are one
linewidth of a mode (at that frequency) beyond the location of the
outer $m = \pm l$ peaks. The dotted and dashed lines do the same,
respectively, for the $l=4$ and 5. The right-hand panel is instead
centred on the $l=1$ (upper triplet) and $l=3$ modes (lower
triplet). Inspection of the plots indicates that the potential for
leakage from higher $l$ is clearly a bigger cause for concern for
the even-degree pairs, because of the closer proximity of the $l=4$
and 5 peaks.

\begin{figure*}
 \centerline{ \epsfbox{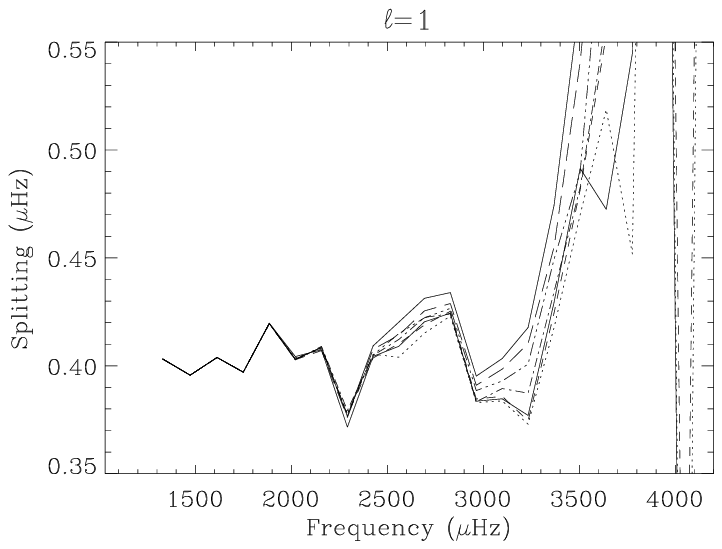}
 \quad \epsfbox{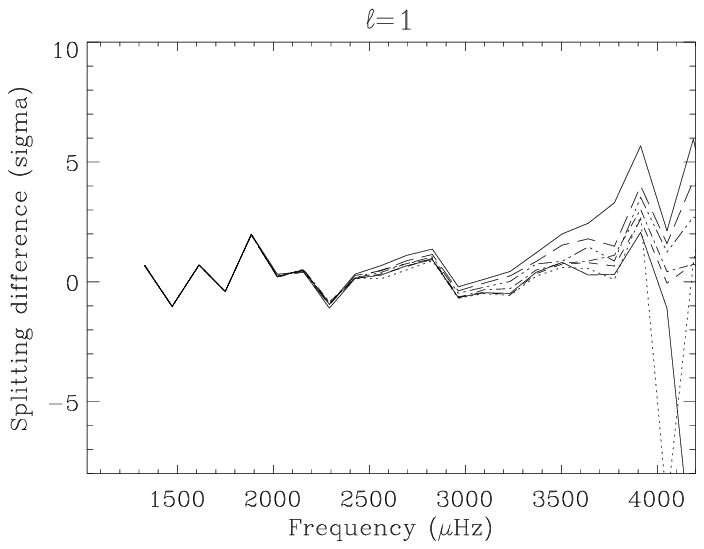}}
 \centerline{ \epsfbox{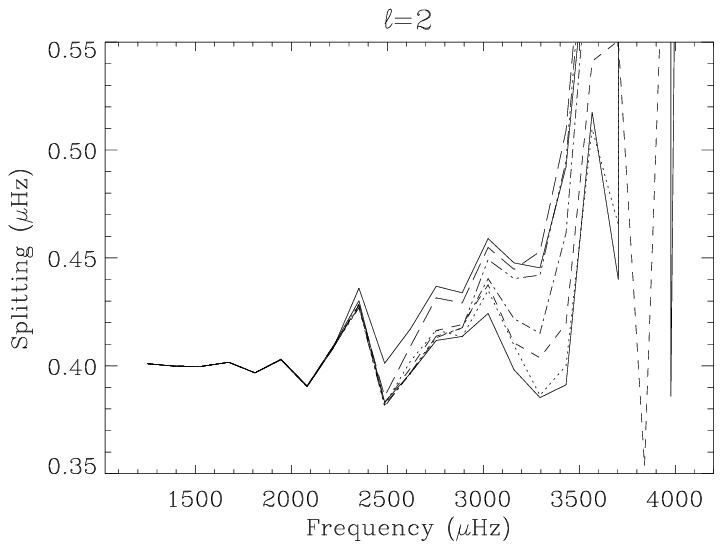}
 \quad \epsfbox{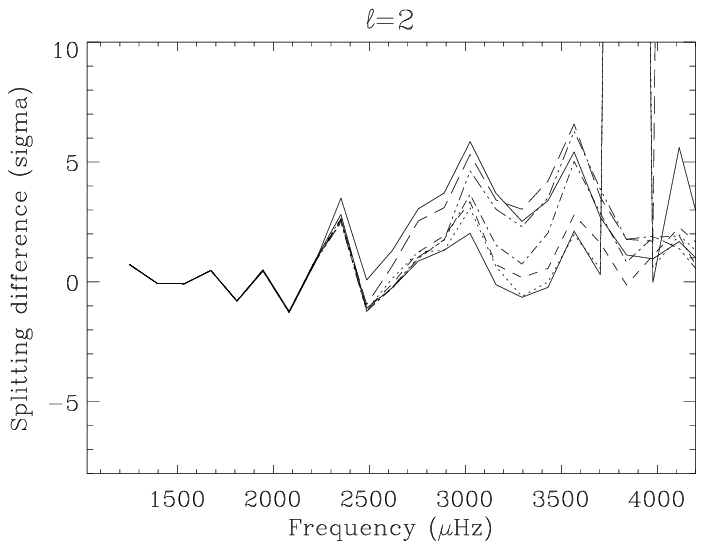}}
 \centerline{ \epsfbox{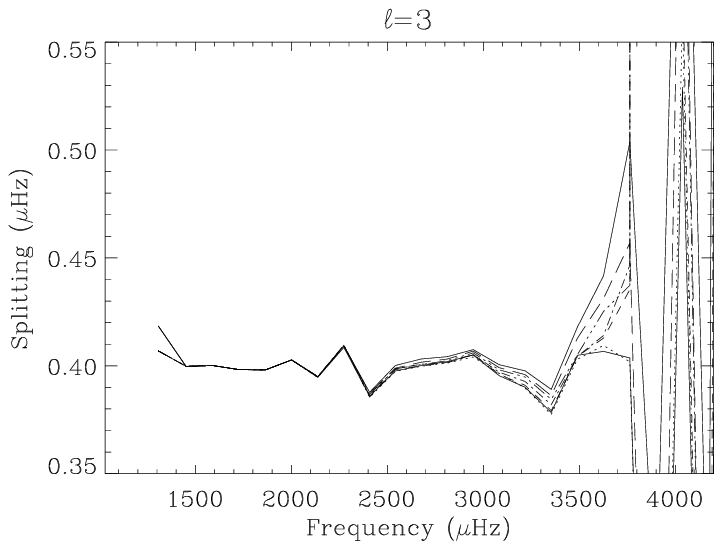}
 \quad \epsfbox{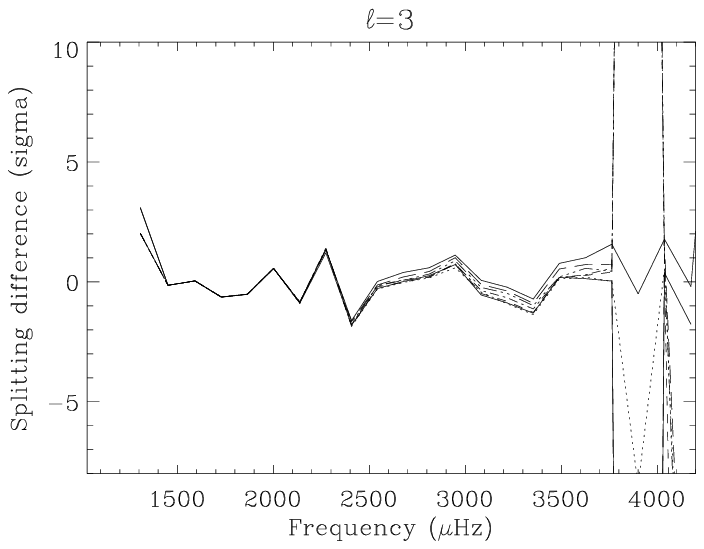}}

\caption{Impact of imposing different sizes of fitting window. The
code of TT was used to generate the results. Window sizes ranged
from 35 (giving data described by lower solid curve in each panel)
up to $65\, \rm \mu Hz$ (uppermost, solid curve in each panel) in
    uniform steps of $5\, \rm \mu Hz$ (the intermediate curves). Shown
in the left-hand panels: the extracted splittings. Shown in the
right-hand panels: fitted minus input splittings, divided by
corresponding mean fitted uncertainties.}

 \label{fig:range}
 \end{figure*}

Fig.~\ref{fig:range} reveals the impact on the extracted splittings
of changing the fitting window size. TT performed fits with window
sizes that ranged, in uniform increments of $5\, \rm \mu Hz$, from
35 to $65\, \rm \mu Hz$. These results bear out the simple
prediction from Fig.~\ref{fig:where}. The $l=2$ splittings are
noticeably affected in the range above $\simeq 2500\, \rm \mu Hz$,
where the increased power in the neighbouring $l=4$ and 5 peaks
means the leakage is most severe. In contrast, the odd-pair $l=1$
and 3 splittings show little change. The relative importance for
each degree is similarly reflected by the significance of the dotted
curves in Fig.~\ref{fig:probs}.



The analysis above, and range of fitting windows actually chosen by
the hounds, suggested that while the window-size effect would have
contributed at $l=2$ to the splitting differences, it was probably
not the largest acting; the biggest effect probably came from the
height-ratio selection.

 \section{Discussion}
 \label{sec:disc}

Nine members of the FLAG group (the hounds) analyzed an artificial
3456-d dataset constructed by the hare (WJC). The data were made to
mimic seismic, Sun-as-a-star observations in Doppler velocity. Fits
were made by the hounds to the low-$l$ resonant peaks in a power
spectrum of the full time series, under the favourable conditions
offered by a 100-per-cent duty cycle of simulated
observations. Estimates of the multiplet frequency splittings of the
$l=1$, 2 and 3 modes were returned to the hare for further scrutiny.
The nature of the data are such that these splittings are close to the
sectoral splittings of the modes.

All nine sets of splittings were affected at high frequencies by the
well-known effects of mode blending. This gave rise to an
overestimation in the splittings, which was most severe at $l=1$. It
is at this degree that the outer components are most closely spaced
in frequency, and the associated splitting uncertainties are as a
result also largest for the dipole modes.

Comparison of the splittings revealed differences between the hounds,
and several sources of bias were identified and studied.  This bias
turned out to be of least significance at $l=1$, where the
\textsc{rms} scatter between the splittings of the hounds was but a
small fraction of the typical fitting uncertainties. Departure of the
fitted splittings from the input values was dominated at $l=1$ by
realization noise, and a variety of fitting strategies and imposed
fitting constraints gave similar results.

Scatter between hounds was in contrast more severe at $l=2$ and 3,
where the lower uncertainties meant bias from several sources came
into play. The simpler case to explain was $l=3$. Here, variation in
the $m$-component height ratios used by hounds in their fitting
models was the dominant source of bias. This variation gave rise in
the first sets of submitted splittings to \textsc{rms} scatter
between hounds at a level, over much of the frequency range, of
twice the typical fitting uncertainties (and extreme differences of
more than $6\sigma$). Re-fitting, with the correct height ratios,
all but eliminated these differences.

At $l=2$, differences arose at the \textsc{rms} level of $\simeq
1\sigma$. These differences came from a combination of effects. The
largest was probably from the aforementioned height ratio choice.
Other bias came from the chosen size of fitting window, and from
a-priori choices over fitting strategy (e.g., how to constrain the
fitting widths in a pair).

What are the implications for future analysis of Sun-as-a-star data?
First, accurate modeling of the $m$-component height ratios is vital
to avoid the introduction of potentially significant bias in the
fitted splittings. This can be done either by accurate and precise
modeling of the instrumental response function, or by estimation of
the ratio by fitting. A careful strategy will largely remove any
bias at $l=3$. At $l=2$, the size of the fitting window also
matters. Sensitivity to this can be reduced by making allowance in
the fitting model for the presence, in the fitting window, of power
from nearby $l=4$ and 5 peaks, and `background' from the slowly
decaying tails of the other even and odd-pair modes. With bias from
the height ratio and window effects largely removed, the choice of
how to model widths in a pair contributes only at higher
frequencies.

It is also worth commenting on implications for asteroseismic
analysis on solar analogues. Differences in internal rates of
rotation, convective properties, and angles of inclination offered
by stars all present their own set of fitting problems.
Faster-than-solar rotation can reduce bias compared to the FLAG
case, since peaks are then more widely spaced in frequency and
easier to distinguish. However, clear advantages only accrue if the
widths of mode peaks are not also significantly increased in size.
The strength of damping, and the width of peaks, is largely
dependent on the convection in the star. For example, a main
sequence star which is hotter than the Sun will tend to support more
vigorous convection in a thinner sub-surface zone; other things
being equal, this will lead to heavier damping of the acoustic
modes, and larger widths in the frequency domain.

The angle of inclination, $i$, has a large effect on the observed
pattern of peaks. The extreme cases, of $i \simeq 0$ and $\simeq 90$
degrees (the latter relevant to the extant helioseismic data), give
patterns of peaks that are least `cluttered'. While this can help
reduce bias from blending of adjacent components, the fact that
components are missing means that information on the rotation is
limited. An intermediate angle may result in all components having
reasonable visibility (provided the background noise is small);
however, the spacing between observed components is then approximately
half that in the extreme $i$ cases. The impact of mode blending is
then much more severe.

Our studies indicate that accurate estimation of the height ratios
-- which have a strong dependence on $i$ -- will be important for
accurate asteroseismic inference on rotation. Gizon \& Solanki
(2003) and Ballot et al. (2004) have already looked at this problem
with artificial data. They found it was possible to extract
well-constrained estimates of $i$ from time series of fairly modest
length (5 to 6 months), provided stars rotated at least twice as
fast as the Sun, and $i \ga 30$ degrees. An \textit{astero}-FLAG
investigation on short dataset lengths is clearly desirable to fully
test the conclusions of this paper for realistic asteroseismic
scenarios.

\subsection*{ACKNOWLEDGMENTS}

The NSO/Kitt Peak data, used as part of the FLAG time series
construction, were produced cooperatively by NSF/NOAO, NASA/GSFC and
NOAA/SEL. We thank the referee, S.~Korzennik, for his careful review
of the paper.  The authors acknowledge the significant contributions
to helioseismology made by their recently deceased colleague,
G.~R.~Isaak.

\end{document}